\documentclass{JHEP}
\input epsf
\usepackage{epsfig}
\usepackage{amssymb}
\usepackage[active]{srcltx}

\newcommand{\bea}{\begin{eqnarray}}
\newcommand{\eea}{\end{eqnarray}}
\newcommand{\bean}{\begin{eqnarray*}}
\newcommand{\eean}{\end{eqnarray*}}
\newcommand{\nn}{\nonumber \\}

\def\IR{\mathbb{R}}

\def\O #1{\overline{#1}}

\def\W #1{\widetilde{#1}}

\def\braket#1{\left\langle #1 \right\rangle}

\def\gb #1{ \left\langle #1 \right]}

\def\det{\mathop{\rm det}}

\def\wt{\widetilde}

\def\th{{\theta}}

\def\a{{\alpha}}

\def\b{{\beta}}

\def\la{\lambda}
\def\eps{\epsilon}

\def\vev{\braket}

\def\bvev#1{\left[ #1 \right]}
\def\Spaa{\vev}
\def\Spbb{\bvev}
\def\Spab{\gb}

\def\Label#1{\label{#1}%
  \smash{\hbox to0pt{\raise1ex\hbox{\tiny[#1]}\hss}}}

\preprint{
\\
{\tt hep-th/}}
\title{Cross Section Evaluation by Spinor Integration I: The massless case in 4D}
\author{ Bo Feng$^{\diamond \dagger}$, Rijun Huang$^{\ast}$, Yin Jia$^{\ast}$,  Mingxing Luo$^{\ast}$,  Honghui Wang$^{\ast}$~~~~\\
$^\diamond$Center of Mathematical Science, Zhejiang University, Hangzhou 310027, China\\
$^{\ast}$Zhejiang Institute of Modern Physics, Physics Department, Zhejiang University, Hangzhou 310027, China\\
$^\dagger$ Division of Applied Mathematica and Theoretical Physics, \\
China Institute for Advanced Study, Central University of Finance\\
and Economics, Beijing, 100081, China\\}

\abstract{ To get the total cross section of one interaction from
its amplitude ${\cal M}$, one needs to integrate $|{\cal M}|^2$ over
phase spaces of all out-going particles. Starting from this paper,
we will propose a new method to perform such integrations, which is inspired
by the reduced phase space integration of one-loop unitarity cut
developed in the last few years. The new method reduces one
constrained three-dimension momentum space integration to an
one-dimensional integration, plus one possible Feynman parameter
integration. There is no need to specify a reference framework in
our calculation, since every step is manifestly Lorentz invariant by
the new method. The current paper is the first paper of a series for the new method.
Here we have exclusively focused on  massless particles in 4D.
There is no need to carve out a complicated integration region in
the phase space for this particular simple case because  the
integration region is always simply $[0,1]$. }

\begin{document}

\section{Introduction}

In the last few years there were great progresses in the evaluation
of one-loop amplitudes for general field theory (see
\cite{Bern:2008ef} and references within). One of such progresses is
the unitarity cut method, which was initiated  in
\cite{Bern:1994zx, Bern:1994cg} and then pushed by Witten's
``twistor program" \cite{Witten:2003nn}. One key achievement along
this line is the reduced unitarity phase space integration using
``holomorphic anomaly"\cite{Cachazo:2004by}. More accurately, the
reduced unitarity phase space integration is given by\footnote{The
$\delta^+(~)$ means the integrated region with $l_0\geq 0$.}
\bea \int d^4 L \delta^+(L^2) \delta^+((L-K)^2)G(L)~. \eea
With the first delta-function $\delta^+(L^2)$ we can reduce the
measure $\int d^4L \delta^+(L^2)$ into the integration with variables
$t,\la, \W\la$, where $t$ is a affine variable and $\la, \W\la$ are spinor variables.
With the second delta-function $\delta^+((L-K)^2)$, we can integrate
$t$ out and get
\bea \int \Spaa{\la|d\la}\Spbb{\W\la|d\W\la} f(\la, \W\la)~. \eea
Before the work of  \cite{Cachazo:2004by}, the evaluation of
remaining two-dimensional integration is  a very hard task and
has blocked  practical applications of unitarity cut method.
From \cite{Cachazo:2004by} is that people realized that
the remaining integrations over $\la,\W\la$ can be obtained by
reading out residues of corresponding poles. In another word, there
is no need for integrations  and everything is just algebraic
manipulation.  This method is usually referred to as ``Spinor
Integration Method" or the spinor method for short.

Clearly, the success of the spinor integration method is due to the
presence of two delta-functions. Using this technique,  we are able
to perform any unitarity phase space integration at one-loop  in
pure 4 dimension\cite{Britto:2004nj, Britto:2005ha, Britto:2006sj}.
Originally, spinor is tightly related to null momentum in pure 4D.
However, for  practical applications, it will be useful to
generalize it to general D-dimension as well as massive particles.
This goal has been achieved late in \cite{Anastasiou:2006jv,
Anastasiou:2006gt}.

To get the  cross section,  we need to integrate
the  physical phase space\footnote{To make things clear, we use
unitarity phase space and physical phase space to distinguish these
two cases.} of all (or some) out-going particles,
\bea \label {physical-space}c\int \prod_{f} {d^3 p_f\over 2E_f}
|{\cal M}|^2\delta^4(P_{in}-P_{out})\eea
where $\delta^4(P_{in}-P_{out})$ is the energy-momentum conservation
condition and $c$ is a function depending on the in-coming particles
(for example, $c^{-1}=4 E_A E_B |v_A-v_B|$ in $2\to n$ progresses).
For simplicity, we  omit $c$ in this paper.

Usually, integrations as such are difficult to perform, especially
when there are many out-going particles. For LHC experiments,
channels with four or even five out-going particles play important
role in  search of new physics
(see, for example, \cite{GehrmannDe Ridder:2003bm} and references in this
paper). Aiming at this task,  we will try to do  the physical  phase
space integration using the spinor integration method.

The advantage of the spinor method is that it  reduces the
constrained three-dimension momentum space integration to an
one-dimensional integration, plus one possible Feynman parameter
integration. All remaining integrations are scalar type, i.e., the
integrand is manifestly Lorentz invariant, so there is no need to
specify a reference framework in our calculation. Furthermore the
 integration region in the phase space also becomes simpler.

The current paper is the first of a serys of work we will take to
complete our new method. In this paper, we will focus on massless
particles in 4D. For this case, with familiar infrared/collinear
divergences for massless out-going particles, there will be infinity
after the phase space integration. To get sensible physical
quantities, some regularization is needed.  Due to this
difficulty,  the result in this paper is still far from the practical
applications. It is the basis for all late work. We will
deal with massive particles in the second paper and general D-dimension integration in the third part.

For massless particles in pure 4D as focused on in this paper, some
simplifications happen. For example, the integration region in phase
space will always be $[0,1]$ in our method. This will be
modified to be nontrivial functions of mass and energy when the
particles are massive as to be presented in second paper.

The main aim of this paper is to laid out the framework of our
new method, so most examples in this paper are not for real cross
sections. These examples is to demonstrate the salient character
of spinor integration, such as frame independence and simple integration region.

The outline of this paper is the following. In section 2, we
transfer the physical phase space integrations into the form of
spinor integrations. We start with the case when there are two and
three outgoing particles, then generalize to cases with $n\geq 3$
recursively. The recursive feature (not recursion relation) is one
of advantages of the spinor integration method.

In the following three sections, our method has been demonstrated
with  simple examples with  two, three and four outgoing particles,
respectively. There will also a brief discussion on the IR/collinear
divergence problem related to massless particles in section 3.

In section 6, we summarize our results along with general
discussions and remarks.

There are two appendixes. In Appendix A the evaluation of pure four
particle phase space directly using momentum components has been
given as  to compare with spinor integration method presented in
main text. In Appendix B we have explained the unfamiliar spinor
integration method from the point of view of integration in complex
plane. From this point, the correctness and the power of this method
becomes obvious.

\subsection{The reduced phase space integration of unitarity cut}

As we have emphasized above, the key gradient of spinor method is
the existence of two delta-functions. There is one situation where these
two delta-functions arise naturally: it is the reduced phase space
integration of one-loop unitarity cut.
In this subsection, we  review how to perform
the phase space integration of one-loop unitarity cut by spinor method.

\begin{figure}
\begin{center}
  \includegraphics[height=4cm]{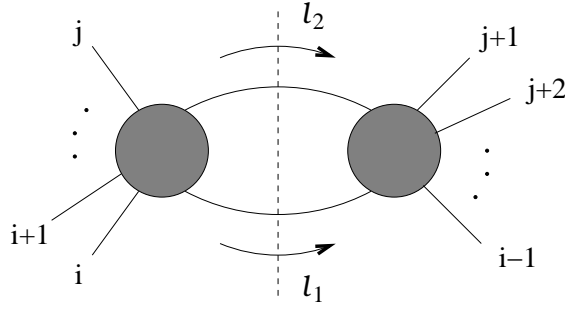}\\
  \caption{Representation of the cut integral. Left and right legs are on-shell. Internal lines represent
  the cut propagators.}\label{cut-pic}
  \end{center}
\end{figure}

{\bf Unitarity Cut}:
 Consider the unitarity cut in the
$(i,i+1,\ldots,j-1,j)$-channel, as shown in fig. \ref{cut-pic}. The
cut integral is 
\bea &&C_{i,i+1,\ldots,j-1,j}\nn &=&  \int d\mu
A^{\mathrm{tree}}(l_1,i,i+1,\ldots,j-1,j,l_2)A^{\mathrm{tree}}
((-l_2),j+1,j+2,\ldots,i-2,i-1,(-l_1))\label{u-cut},\eea 
where $d\mu = d^4 l_1d^4
l_2\delta^{(+)}(l_1^2)\delta^{(+)}(l_2^2)\delta^4(l_1+l_2-P_{ij})$
is the Lorentz invariant phase space measure of two light-like
vectors $(l_1,l_2)$ constrained by the energy-momentum conservation,
and $P_{ij} = p_i + \ldots + p_j$. Notice that the integrand is
similar to $|{\cal M}|^2$ in eq (\ref{physical-space}). So we can
make full use of this point in our calculations.

Using the $\delta^4(~)$ function to perform the $l_2$ integration
gives
\bea C_{i,i+1,\ldots,j-1,j} &=&\int d^4
l_1\delta^{(+)}(l_1^2)\delta^{(+)}((l_1-P_{ij})^2)\nn*
 && \times A^{\mathrm{tree}}(l_1,i,i+1,\ldots,j-1,j,l_2)A^{\mathrm{tree}}
(-l_2,j+1,j+2,\ldots,i-2,i-1,-l_1)\eea
Note that the Lorentz invariant measure of a null vector $l$ can be
represented as a measure over $\IR^+\times \mathbb{CP}^1\times
\mathbb{CP}^1$. The contour of integration is a certain diagonal
$\Bbb{CP}^1$. Explicitly, one writes $l_{a\dot a} =
t\lambda_a\tilde\lambda_{\dot a}$, and then\footnote{Here we have
used the QCD convention so measure is $\vev{\la ~d\la }[d\wt{\la} ~
\wt{\la} ]$ instead of the twistor convention $\vev{\la ~d\la
}[\wt{\la} ~ d\wt{\la} ]$.}
\bea  \label{change}\int d^4l \delta^{(+)}(l^2) ( \bullet ) \sim
\int_0^{\infty }t\; dt \int_{\wt\lambda = \bar\lambda} \vev{\la
~d\la }[d\wt{\la} ~ \wt{\la} ] (\bullet),\eea
where $(\bullet )$ represents a generic integrand up to an overall
numerical factor. So we have\footnote{Where we have written
$(\ell_1-P_{ij})^2= P_{ij}^2-t\gb{\la|P_{ij}|\W\la}$, thus we have
used the QCD convention instead of the twistor convention.}
\bea \label{u-cut-1}C_{i,\ldots,j}\sim \int_0^{\infty }t\; dt \int
(-)\Spaa{\la~d\la}\Spbb{\W\la~d\W\la}
\delta^{(+)}\left(P_{ij}^2-t\Spab{\la|P_{ij}|\W\la}\right)G(\la,\wt\la,t).
\eea
$G(\la,\wt\la,t)$ arises from the product of the two tree-level
amplitudes in (\ref{u-cut}). In order to get $G(\la,\wt\la,t)$ in
actual calculations, we have to write expressions of the form
$\Spaa{\bullet,l_2}$ or $\Spbb{\bullet,l_2}$ in terms of $l=l_1$
($l_1$ is always substituted by $l$ when there is no possibility of
confusion). A systematic way of doing this is as follows:
\bea \Spaa{\bullet,l_2} = {\Spaa{\bullet,l_2}\Spbb{l_2 ~ l_1}\over
\Spbb{l_2 ~ l_1} }= {\Spab{\bullet|l_2|l_1}\over \Spbb{l_2 ~
l_1}}={\Spab{\bullet|P_{ij}|l_1}\over [l_2~l_1]}.\eea
A similar identity is valid for $\Spbb{\bullet,l_2}$. The factors
$\Spbb{l_2 ~ l_1}$ and $\Spaa{l_2 ~ l_1}$ all pair up in the end
allowing for the use of the vector form of $l_2$. This is because
the product of amplitudes must be invariant under the scaling
$z\la_{l_2} $ and $z^{-1}\wt\la_{l_2} $. For (\ref{u-cut-1}),
integrating over $t$ yields
\bea \label{u-cut-2}C_{i,\ldots,j} \sim
P_{ij}^2\int{\Spaa{\la~d\la}\Spbb{\W\la~d\W\la}\over
\Spab{\la|P_{ij}|\W\la}^2} G\left(\la,\wt\la,{P_{ij}^2\over
\Spab{\la|P_{ij}|\W\la}}\right).\eea
The requirement of degree zero in $\wt\la$ implies that $G$ in
(\ref{u-cut-2}) can be written as a sum of terms in the
form\footnote{It is worth to emphasize that this general form is
only true for the one-loop calculation. When we apply our method to
cross section evaluation, we will have other forms of input. The way
to deal with them generally is given in Appendix B.}

\bea {\prod_{i=1}^k \Spab{\la|R_i|\W\la}\over
\Spab{\la|P_{ij}|\W\la}^{a}\prod_{j=1}^{k-a}
\Spab{\la|Q_j|\W\la}},\eea
where $R_i, Q_j$ are functions of external momenta.

{\bf Canonical Splitting}: In order to calculate the integral
efficiently,  we can reduce the integrand by separating the
denominator factors with $\wt\la$ as much as possible, at the cost
of more terms. When there is a product $[a~\W\la][b~\W\la]$ in the
denominator, multiply both numerator and denominator by $[a~b]$.
Applying Schouten's identity yields
\bea
\Spbb{i~j}\Spbb{k~l}=\Spbb{i~k}\Spbb{j~l}+\Spbb{i~l}\Spbb{k~j}\eea
in the numerator with another factor $[c~\W\la]$ (which must exist
by homogeneity when the degree of $\W\la$ in denominator is equal to
or more than three). Thus we get two terms with $[a~\W\la]$ or
$[b~\W\la]$ in the numerator, canceling one of the denominator
factors. The result, in terms of $\wt\la$, is a denominator of the
form $\prod_r \Spab{\la|Q_r|\W\la}\Spbb{A~\W\la}$ in every term.
Similarly $\Spab{\la|Q_r|\W\la}$ can be treated by writing
$\Spab{\la|Q_r|\W\la}=\Spbb{\wt Q_r~\W\la}$, where
$\wt\la_{Q_r}=-(Q_r)_{a\dot a}\la_l^a$.

Using this procedure repeatedly and noticing the integrand has a
degree of $-2$ in $\wt\la$, we end up with two kinds of possible
integrals:
\bea \mathcal{I}_A=-\int{\Spaa{\la~d\la}\Spbb{\wt\la~d\wt\la}
\prod_{i=1}^k \Spab{\la|R_i|\W\la}\over
\Spab{\la|P_{ij}|\W\la}^{2+k}};\qquad
\mathcal{I}_B=-\int{\Spaa{\la~d\la}\Spbb{\wt\la~d\wt\la}\over
\Spab{\la|P_{ij}|\W\la}\Spab{\la|Q_r|\W\la}}.\eea
$\mathcal{I}_B$ can be changed to
\bea \mathcal{I}_B=-\int_0^1 dx
\int{\Spaa{\la~d\la}\Spbb{\wt\la~d\wt\la}\over
\Spab{\la|xP_{ij}+(1-x) Q_r|\W\la}^2},~~~\label{I-B}\eea
which reduces to $\mathcal{I}_A$ up to a Feynman parametrization
integration. On the other hand, $\mathcal{I}_A$ is just the
$\prod_{i} y_i$ component of an auxiliary integration\footnote{There
are other ways to do it directly without using the auxiliary
integration. See reference \cite{Britto:2005ha}.}
\bea \mathcal{I}_{aux}=\int{\Spaa{\la~d\la}\Spbb{\wt\la~d\wt\la}
 \Spab{\la|R|\W\la}^k\over
\Spab{\la|P_{ij}|\W\la}^{2+k}},~~~~R=\sum_{i} y_i
R_i,~~~~\label{Inte-R}\eea

Now all integrations are reduced into the one given by
$\mathcal{I}_{aux}$. It can in turn be written as a total derivative
by using
\bea { [\wt\la~d\wt\la] [\eta~\wt\la]^n\over
\gb{\la|P|\wt\la}^{n+2}} = [d\wt\la~\partial_{\wt\la}] \left(
{1\over (n+1) \gb{\la|P|\eta}} { [\eta~\wt\la]^{n+1} \over
\gb{\la|P|\wt\la}^{n+1}} \right) \label{GETPOLE}. \eea
Thus, the evaluation of formula (\ref{Inte-R}) is transformed into
the reading of residues of
\bean {1\over (k+1) \Spaa{\la|R|P_{ij}|\la}} {
\Spab{\la|R|\wt\la}^{k+1} \over \Spab{\la|P_{ij}|\W\la}^{k+1}} \eean
where we need to sum up two possible poles from the factor
$\Spaa{\la|R|P_{ij}|\la}$. The way to sum up these two poles has
been discussed in detail, for example,  in eq (70), (74) of
\cite{Britto:2006fc}. In fact,  the result can be written down
directly as
\bea  \label{res-I_{aux}} {-[( -2P\cdot R+\sqrt{\Delta})^{k+1}-(
-2P\cdot R-\sqrt{\Delta})^{k+1}]\over (k+1) \sqrt{\Delta}
(2P^2)^{k+1}}\eea
with $\Delta=(2P\cdot R)^2-4 P^2 R^2$. It is easy to see that
numerator must be of the form $\sum_a\sqrt{\Delta}^{2a+1} (-2P\cdot
R)^{k-2a}$. The result is a rational function with the highest power
of $R$ to be $k$, as required.

Before we end this part, let us emphasize that above procedure works
for general one-loop calculation, but when we try to do the cross
section calculation, we may meet new kinds of singularities and we
need to generalize above procedure. The generalization has been
discussed in Appendix B in some details. {\sl The
basis idea is still to find formula like the one in (\ref{GETPOLE})
and then take the residues}.

\section{Spinor method for the physical phase space}

In this section we will establish the general framework for the
physical phase space integration using spinor method.
We will focus on massless particles in pure 4D.
For massive particles and in general D-dimension, it will be
discussed subsequent work.

Before going into detail, let us make a simple observation:  for
two out-going particles, the physical phase space integration is the
same as the unitarity phase space integration, which can be done
directly by the spinor integration method. Difficulties arise when
$n\geq 3$. We will first deal with the case $n=3$ and then
generalize to arbitrary out-going particles.

Notice that we
can rewrite
\bea \int {d^3 p_f\over 2E_f}\sim \int d^4 p_f
\delta^+(p_f^2) ,\eea
Then
when there are only two out-going particles, we have
\bea \label{two-particle} \prod_{f=1,2}\int d^4 p_f
\delta^+(p_f^2) \delta^4(P_{in}-p_1-p_2)= \int d^4 p_1
\delta^+(p_1^2)\delta^+((p_1-P_{in})^2)~.\eea
The two delta-functions for $\int d^4 p_1$  is exactly what one needs for
the unitarity phase space integration, thus we can use the spinor integration method to
evaluate the cross section  efficiently.

\subsection{The phase space integration with three out-going particles}
When $n=3$ we have
\bean I &\sim & \prod_{i=1}^3 \int d^4 L_i \delta^+(L_i^2))
\delta^4(K-\sum L_i)f(L_1,L_2, L_3)\\
& = &  \int d^4 L_1 \delta^+(L_1^2)\int d^4L_2
\delta^+(L_2^2)\delta^+ ((K-L_1-L_2)^2)f(L_1,L_2,
K-L_1-L_2)\eean
The integration over $L_2$ with two delta-functions can be done by
the spinor integration method. However, for $\int d^4 L_1$ we have
only one delta-function. The same difficulty also
arises when $n>3$: for the integration over each momentum, we need
another delta function in order to make use of the spinor
integration method. Then the problem is how to rewrite the integration
measure in such a form where one more delta-function shows up. The
solution of this problem is the application of a Faddeev-Popov like
method. We will show how it works in detail soon.

Having the solution for the $n=3$ case, the solution for arbitrary
$n\geq 4$ can be obtained by doing the integration recursively one
by one. Keeping  $(n-1)$-momenta fixed, we can integrate the
$n$-th momentum. After that, we keep  $(n-2)$-momenta fixed and
integrate the $(n-1)$-th momentum and so on until $n=3$.

Now let us do the case $n=3$ explicitly. The integration
over $L_2$ is trivial. We will denote the result after this
integration as $\W f(L_1)$. Then
\bea \label{typical}I \sim \int d^4 L_1 \delta^+(L_1^2)\W
f(L_1)\eea
which is a typical phase space integration and we rewrite it as
\bea I=\int {d^4 \ell\over (2\pi)^3} \delta^+(\ell^2)\W
f(\ell).\eea

To get the standard form suitable for spinor integration method, we
need to insert another $\delta$-function. To do so, let us consider
the kinematics in more detail. The total energy-momentum tensor of
out-going particles is given by $K$, which has following properties:
$K_0\geq 0$ for positive energy component and $K^2\geq 0$.
With these in mind, let us consider following expression
\bea I_z\equiv \int dz \delta( (zK- \ell)^2) & = & \int dz \delta (
z^2 K^2- z(2K\cdot \ell)+\ell^2)=\int dz \sum_{i=1,2}
{\delta(z-z_i)\over |2 z_i K^2- (2K\cdot
\ell)|},~~~~\label{I-z-def}\eea
where
\bea z_i & = & { (2K\cdot \ell)\pm \sqrt{(2K\cdot \ell)^2-4 K^2
\ell^2}\over 2 K^2}.\eea
Let's determine the integration region of $z$ for the physical system.
In the center-of-mass frame, $K=(E,0,0,0)$, the out-going momentum $\ell=(E_1, P,0,0)$ and $E_1^2= P^2$.
Conservation laws ensure that the energy-momentum of all other
out-going particles to be $K_{other}=(E-E_1,-P,0,0)$.
For physical
particles we have $K_{other}^2\geq 0$, i.e., $(E-E_1)^2\geq P^2$, or
$E\geq E_1+|P|$.
%
%
%
%
%
Then we get
\bea z_{i}= {2 E(E_1\pm P)\over 2 E^2}={ E_1\pm P\over E}.\eea
When $m=0$, we have $E_1=|P|$ and especially $z_{-}=0$. Note that $z=0$ is always a solution when we combine the massless condition $\ell^2=0$  in (\ref{I-z-def}). Thus to 
have a physical meaningful solution, we can keep only $z_+$
solution.  It is easy to see that \bea 0\leq z_+\leq
1~~~~~\label{z-reg}.\eea
That is, $I_z$ defined in (\ref{I-z-def}) makes sense
physically only in region $z\in [0,1]$. With this consideration, $I_z$ can be defined 
\bea I_z & = & \int_0^1 dz {\delta(z-z_+)\over
|\sqrt{(2K\cdot \ell)^2-4 K^2 \ell^2}|}={1\over |\sqrt{(2K\cdot
\ell)^2-4 K^2 \ell^2}|}.\eea
In another word we have
\bea {|\sqrt{(2K\cdot \ell)^2-4 K^2 \ell^2}|}\int_0^1 dz
\delta( (zK- \ell)^2) =1.\eea
In the same spirit of the Faddeev-Popov method in gauge-fixings, we insert it into eq
(\ref{typical})
\bean I & = &{1\over (2\pi)^3} \int d^4 \ell \delta^+(\ell^2)
{|\sqrt{(2K\cdot \ell)^2-4 K^2 \ell^2}|}\int_0^1 dz \delta(
(zK- \ell)^2) \W f(\ell)\\
%
%
& = & {1\over (2\pi)^3}\int_0^1 dz\int d^4 \ell
\delta^+(\ell^2){(2K\cdot \ell)} \delta( (zK- \ell)^2)\W
f(\ell).\eean
By use of eq (\ref{change}) and taking the QCD convention $2a\cdot
b=\Spab{a|b|a}$, then
\bean I
%
& = & {-c\over (2\pi)^3}\int_0^1 dz\int
\Spaa{\la|d\la}\Spbb{\W\la|d\W\la}\int t dt {zK^2}\delta\left( z^2 K^2- z t \Spab{\la|K|\W\la}\right)\W f(\la,\W\la,t)\\
%
%
& = & {-c\over (2\pi)^3}\int_0^1 dz {z}\int
\Spaa{\la|d\la}\Spbb{\W\la|d\W\la} \int dt \delta\left( {z
K^2\over\Spab{\la|K|\W\la}} -  t \right)\left( {
K^2\over\Spab{\la|K|\W\la}}\right)^2\W f(\la,\W\la,t).\eean

Putting all together we have
\bea I & = & \int {d^4 \ell\over (2\pi)^3} \delta^+(\ell^2)\W
f(\ell)\nonumber \\
&= &  {-c\over (2\pi)^3}\int_0^1 dz {z}\int
\Spaa{\la|d\la}\Spbb{\W\la|d\W\la} \int dt \delta\left( {z
K^2\over\Spab{\la|K|\W\la}} -  t \right)\left( {
K^2\over\Spab{\la|K|\W\la}}\right)^2\W
f(\la,\W\la,t).~~~~\label{I-phase}\eea
where $c={\pi /2}$ is related to the Jacobian of changing integration
variables and the way we have taken the residues, in which we will omit the $2\pi i$ factor.
When the integrand is of the form in $\mathcal{I}_A$, eq. (\ref{I-phase}) is all one needs to evaluate.
When the integrand is of the form in  $\mathcal{I}_B$,
one needs one more Feynman parametrization to put it in the form of $\mathcal{I}_A$,
as discussed in section 1.1.

Now we make two remarks. The first is about the choice of momentum $K$. It is not arbitrary: the
$K$ must the the total energy-momentum tensor of all
would-be-integrated momenta. The second is the integration region $z\in [0,1]$. 
In fact, the discussion of region of $z$ is nothing, but the discussion of integration boundary
faced by the evaluation of cross section. The particular simple
result $[0,1]$ is special for massless particles in 4D. If it
is massive, the region of $z$ will be function of mass and
total energy-momentum tensor $K$.

\subsection{Arbitrary number of out-going particles and the recursive method}

Having worked out the processes of three out-going
particles,  we can do the integration for arbitrary $n$ out-going
particles recursively
\bea \label{recusion}I_n(f;K; P_j) &\equiv&{1\over (2\pi)^{3n}} \int
{d^3L_n \over 2E_n} \int {d^3L_{n-1} \over 2E_{n-1}} \cdots \int
{d^3L_1 \over 2E_1} (2\pi)^4\delta^4(K-\sum_{i=1}^nL_i)
f(L_n,L_{n-1},\ldots,L_1)\nn
 &=& {1\over (2\pi)^3}\int {d^3L_n \over 2E_n} I_{n-1}(f;K-L_n; L_n, P_j),\quad (n\geqslant 3).\eea
where in the second line the idea is showing explicitly. More
accurately, assuming $f$ is the function of $n$ should-be-integrated
momenta $L_i$, as well as the total energy-momentum tensor
$K=\sum_{i=1}^n L_i$ and other external momenta $P_j$, the
integration can be done  in two steps. At the first step, we leave
the $L_n$ un-integrated, i.e., we integrate other $(n-1)$ momenta
$L_i$. For this case, the total energy-momentum tensor should be
$(K-L_n)$. After the first step, we integrate the left $L_n$ again
using  (\ref{I-phase}) but now with the total energy-momentum tensor
$K$. Again, there is no need to carve
out a complicated integration region in the phase space for massless
particles. The integrations are always simply over the interval
$[0,1]$.

The idea of recursive evaluation is very natural and has been
applied in other methods, for example,  straightforward
evaluation using  momentum components. This direct evaluation will
be simplified if we choose right reference frame, for example, the
center-of-mass frame. In such case, it usually takes efforts to
rewrite results after integration as the Lorentz invariant form.
This becomes more severe when we have multiple out-going particles.

In contrast, when the spinor integration method is used recursively,
Lorentz invariant expressions are obtained automatically at each step.
There is no need to specify any frame and this simplifies the calculation process greatly.

Finally we want to make some observations. First, for each $\int
d^3p$ integration, we have one $\int dz$ integration, thus for $n$
out-going particles, we have $(n-2)$ $z$-integrations (there maybe
other Feynman parameter integrations from spinor integrations). When
$(n-2)$ is large, usually we can not find the analytic expressions,
but since for each $z$ the integration region is $[0,1]$, the
numerical evaluation should be easy to realize.

Secondly, When converting from momentum variables to spinor
variables in all integrations, there will be Jacobi factors floating
around. In addition, when taking the residues of poles, we have
omitted the $2\pi i$ factor. That is to say, normalization factors
will be needed in the conversion. In eq. (\ref{I-phase}), it is
$c=\pi/2$.
Fortunately, for given number of out-going particles, this normalization factor is universal.
The simplest way to fix these normalization
factors is to calculate the pure physical phase space volume by both
methods, as to be shown in the following examples.

\section{Example one: two out-going particles}

Let us start with the simplest case of two out-going particles. As
we have mentioned in introduction, it has exactly the suitable form
for the spinor integration method. In the following three
subsections, we will use the spinor method to perform the
integration. A brief review of method has been presented in the
introduction and more details can be found in references
\cite{Britto:2005ha}-\cite{Britto:2006fc}. For comparison, we also
list the results from standard momentum integration. In the final
subsection, we will briefly discuss the IR/collinear divergences.

As we have mentioned in the introduction, although these examples
are very simple and bear little practical importance, they are good
for  the demonstration of our method, from which the character of
our method is clear. The validity of our method for
general input  shows in the Appendix B.

\subsection{The pure phase space integration}

 We
denote the physical phase space integration of $n$ out-going
particles as $I_n^{s\, \mathrm{or} \, m}(f;K)$, where $s$ stands for
the spinor method and $m$ the momentum method. The $K$ is the sum of
momenta of these $n$ particles and $f$ can be a function of $K$ and
other external variables as well as the $n$ would-be-integrated
momenta.

{\bf Spinor integration method} : The integration is given by
\bean I_2^s(1;K) &=& \int {d^4L_2\over (2\pi)^3} {d^4L_1\over
(2\pi)^3}\delta^+(L_2^2)\delta^+(L_1^2)(2\pi)^4\delta^4(K-L_2-L_1).\\
& = & {-\W c\over (2\pi)^2} \int tdt
\int\Spaa{\la|d\la}\Spbb{\W\la|d\W\la} \delta((K-L)^2)\eean
where $\W c={\pi/ 2}$ is the Jacobi factor of changing integration variables.
As mentioned at the end of the last section, we will omit the $2\pi i$ factor when taking the residues of poles.
It has been calculated in various references \cite{Britto:2005ha} and the result is
\bea  I_2^s(1;K)&=&{1\over(2\pi)^2}{\pi\over 2}.\eea
%
%
%
%
%
%

{\bf Momentum integration method}:
The expression is given by
\bea I_2^m(1;K) &=&\prod_{i=1,2} \int
\frac{dL_i^3}{(2\pi)^32E_i}(2\pi)^4\delta^4(K-L_2-L_1).\eea
Taking  the center-of-mass frame, where $K=(E,0,0,0)$, and by use of
%
\bean {\delta(E-E_1-E_2) \over 2E_1 }=
\delta((E-E_2)^2-E_1^2)=\delta((K-L_2-L_1)^2)=\delta(E^2-2EE_2).\eean
 we get
\bea I_2^m(1;K) ={1\over(2\pi)^2} \int d\Omega \int
{E_2^2dE_2\over2E_2}\delta(E^2-2EE_2)
&=&{1\over(2\pi)^2}{\pi\over2},\eea
which is identical to the result obtained from the spinor method. It
is worth to notice that a suitable choice of reference frame has been
made to simplify the calculation.


\subsection{The example with $f=2L_1\cdot L_2$}


{\bf Spinor integration method}: With this integrand, we
have\footnote{In fact, using $2L_1\cdot L_2=(L_1+L_2)^2=K^2$, this
example is same as the pure phase space integration. Here we do it
using different way to demonstrate the technique of spinor
integration.}
\bean I_2^s(u;K) &=&\int {d^4L_2\over (2\pi)^3}
{d^4L_1\over (2\pi)^3}\delta^+(L_2^2)\delta^+(L_1^2)(2\pi)^4\delta^4(K-L_2-L_1)2L_1L_2\\
&=& {1\over(2\pi)^2}\int
d^4L_2\delta^+(L_2^2)\delta^+((K-L_2)^2)2L_2(K-L_2). \eean
Using $L_2^2=0$ there is only one term left. Following the process
in the first section, we use (\ref{change}) to rewrite the
integration and integrate $t$ with delta function, then use
(\ref{GETPOLE}) to get the final result
\bea I_2^s(u;K)=\frac{1}{(2\pi)^2}{\pi\over2}K^2. \eea

 {\bf Momentum integration method}: Similar to
the pure phase space integration, we have
\bea I_2^m(u;K)
 &=&
{1\over(2\pi)^2}{\pi\over2}E^2= {1\over(2\pi)^2}{\pi\over2}K^2\eea
which is identical to $I_2^s(u;K)$.



\subsection{The example with $f= {1/(2P_1\cdot
L_2+m^2)}$}
This example is actually of certain practical value. In real cross
section calculations, we meet infrared/collinear divergences when
there are massless particles. One way to regularize such divergences
is to add a mass term in the propagator.

{\bf Spinor integration method}: The integral is
\bean I_2^s(f;K)
&=& {1\over(2\pi)^2}\int
d^4L_2\delta^+(L_2^2)\delta^+((K-L_2)^2){1\over 2P_1\cdot L_2+m^2}
.\eean
 Using (\ref{change}) to rewrite
the integration and integrating $t$ with delta function, we have
\bea I_2^s(f;K)={-1\over(2\pi)^2}{\pi\over 2}\int
\Spaa{\la|d\la}\Spbb{\W\la|d\W\la}{1\over
\Spab{\la|K|\W\la}\Spab{\la|P_1+\a^2K|\W\la}},\eea
where $\a^2={m^2/ E^2}$ and the normalization factor ${\pi/ 2}$ has
been inserted with the spinor integration variables. Introducing a
Feynman parameter, we rewrite the above formula as
\bea {-1\over(2\pi)^2}{\pi\over 2}\int_0^1dx\int
\Spaa{\la|d\la}\Spbb{\W\la|d\W\la}{1\over
\Spab{\la|(1-x)K+x(P_1+\a^2K)|\W\la}^2}.\eea
The result can be written down directly
\bea I_2^s(f;K)={1\over (2\pi)^2}{\pi\over 2}\int_0^1dx{1\over
((1-x)K+x(P_1+\a^2K))^2}= {\pi\over 2}{1\over (2\pi)^2}{1\over
K^2}\ln{(1+{1\over \a^2})}.\eea
%
%
%

{\bf Momentum integration method}: Choosing the center-of-mass frame
with $K=P_1+P_2=L_1+L_2$,  we have following components
$K=(E,0,0,0), P_1=(E/2,0,0,P),P_2=(E/2,0,0,-P)$ and
$L_1=(E_1,-\vec{k_1}),L_2=(E_2,\vec{k_1})$, thus
 $2P_1L_2=EE_2(1-\cos{\th})$, $\th$ is the angle between $P_1$
and $L_2$. Doing the integration we are left with
\bea I_2^m(f;K)=\frac{1}{8(2\pi)}\int_0^{\pi}\sin{\th}d\theta
\frac{1}{{E^2\over 2}(1-\cos{\th})+m^2}=\frac{1}{(2\pi)}{1\over
4E^2}\ln{(1+{1\over \alpha^2})}. \eea
%
Notice that the result has a logarithm-dependence divergence when
$m\to 0$, which is the usual IR divergence related to massless
particles. Again in this calculation, special reference frame has
been chosen while in spinor method, each middle step is Lorentz
invariant.

\subsection{The divergent behavior}

To discuss further the divergent behaviors of cross-sections for massless
particles, we now consider a simple physical process, in which two gluons are
scattered another two $P_1+P_2\to P_3+P_4$ of different helicity
configurations. We will focus on the divergent behaviors in these
calculations.

Given color structure, there are three amplitudes with different
helicity configurations. Two of them are
\bea A_{12}(1^- 2^- 3^+ 4^
+)=\frac{\Spaa{1~2}^3}{\Spaa{2~3}\Spaa{3~4}\Spaa{4~1}},\quad
A_{13}(1^- 2^+ 3^- 4^
+)=\frac{\Spaa{1~3}^4}{\Spaa{1~2}\Spaa{2~3}\Spaa{3~4}\Spaa{4~1}}
.\eea
First consider $A_{12}$ and denote $S_{ij}=(P_i+P_j)^2$.
In the
QCD convention we have\footnote{It is just one term of the real
cross section since we have not added up all color-ordered
contribution.}
\bea f_{12}=|A_{12}|^2= \frac{S_{12}^2}{S_{23}^2}\eea
By the
momentum method, we have
\bean I_2^m(f_{12};K) &=&\int{\frac{d^3 P_3}{(2\pi)^3}\frac{d^3
P_4}{(2\pi)^3}\frac{1}{2E_3}\frac{1}{2E_4}(2\pi)^4
\delta^4(K-P_3-P_4)\frac{(2P_1P_2)^2}{(2P_2P_3)^2}}.\eean
Taking the center-of-mass frame and setting the $z$ axis along the
direction of $\vec{p_1}$, we have $P_1=(E,0,0,P),P_2=(E,0,0,-P)$,
thus the integration is reduced to
\bea I_2^m(f_{12};K)
&=&\frac{1}{2\pi}\int_0^{\pi}\frac{\sin{\theta}d\theta}{2(1-\cos{\theta})^2}.\eea
A singularity appears when  $\vec{p_3}$ is parallel to $\vec{p_2}$
and $\th=0$. This is the familiar collinear singularity.

The same divergent behavior can be observed in the spinor method.
After some elementary algebra, we get the following integration form
\bea I_2^s(f_{12};K)=-\frac{1}{(2\pi)^2}{\pi\over 2}\int
\Spaa{\la|d\la}\Spbb{\W\la|d\W\la}
{S_{12}\over\Spab{\la|P_2|\W\la}^2}\sim {1\over P_2^2}. \eea
Just as we have seen from the momentum method,
a divergence appears,  since  $P_2^2=0$.

The situation for $A_{13}$ is a little more complex. We have
\bean f_{13}=|A_{13}|^2=\frac{S_{13}^4}{S_{12}^2S_{23}^2}.\eean
The momentum method gives
\bean
I_2^m(f_{13};K)&=&\frac{1}{2\pi}\int_0^{\pi}\frac{1}{32}\frac{(1+\cos{\theta})^4}
{(1-\cos{\theta})^2}\sin{\theta}d\theta. \eean It can be rewritten
as \bea
\label{2-3-m}I_2^m(f_{13};K)=\frac{1}{2\pi}\int_0^{\pi}\sin{\theta}d\theta
\left(\frac{1}{2(1-\cos{\theta})^2}-\frac{1}{1-\cos{\theta}}\right)
+\frac{3}{2(2\pi)}-\frac{1}{2(2\pi)}+\frac{1}{12(2\pi)}.\eea
where we have left the first two infinite terms un-integrated.

Now take the spinor method. One has
\bea I_2^s(f_{13};K)& = & -\frac{1}{(2\pi)^2}\int
\Spaa{\la|d\la}\Spbb{\W\la|d\W\la}
\frac{S_{12}\Spab{\la|1|\W\la}^4}{\Spab{\la|2|\W\la}^2\Spab{\la|K|\W\la}^4}.\nonumber\\
& = & -\frac{1}{(2\pi)^2}{\pi\over 2}\int
\Spaa{\la|d\la}\Spbb{\W\la|d\W\la}\left(
\frac{S_{12}}{\Spab{\la|2|\W\la}^2}
-\frac{4S_{12}}{\Spab{\la|2|\W\la}\Spab{\la|K|\W\la}}\right.\nonumber
\\
&&\left.+\frac{6S_{12}}{\Spab{\la|K|\W\la}^2}
-\frac{4S_{12}\Spab{\la|2|\W\la}}{\Spab{\la|K|\W\la}^3}
+\frac{S_{12}\Spab{\la|2|\W\la}^2}{\Spab{\la|K|\W\la}^4}\right).\eea
where we have used the spinor algebra to split the first line into
the second line.

It is not difficult to see that the five terms in
$I_2^s(f_{13};K)$ and $I_2^m(f_{13};K)$ correspond to each other exactly.
In particular, the divergence of $1/(1-\cos\theta)$ is related to the divergence of $1/\Spab{\la|2|\W\la}$.

\section{Example two: three out-going particles}

\subsection{The pure phase space integration}

Let us start from the pure phase space integration and identify the
normalization factor $c=\pi/2$, as mentioned in section 2.1.
From eqs (2.3.20) and (2.3.38) of \cite{Field}, the phase space integration is reduced to
\bea \label{i3m1}I_3^{m}(1;K)={Q^2\over 16 (2\pi)^3} \int_0^1 dx_1
\int_{1-x_1}^1 dx_2={Q^2\over 32 (2\pi)^3}.\eea
where $K^2=Q^2$. In this formula, Lorentz invariant variables $x_1,
x_2$ has been used by initial simplification. This is possible
because for small number of particles, there are only a few Lorentz
invariant quantities we can construct. With more and more particles,
the number of such quantities will increase dramatically.

To make use of the spinor method, we have the integration
\bea I_3^{s}(1;K)  = \int {d^4L_1\over (2\pi)^3} {d^4L_2\over
(2\pi)^3} {d^4L_3\over (2\pi)^3}
\delta^+(L_1^2)\delta^+(L_2^2)\delta^+(L_3^2)(2\pi)^4\delta^4(K-L_1-L_2-L_3).\eea
The $\int
d^4L_2d^4L_3\delta^+(L_2^2)\delta^+(L_3^2)(2\pi)^4\delta^4((K-L_1)-L_2-L_3)$
is just the form of the phase space integration of two outgoing
particles, which we know the result is ${1/ 8\pi}$. Putting it back
we are left only with
\bea {\pi\over 2}\int {d^4 L_1\over (2\pi)^5}\delta^+(L_1^2) & = &
{\pi\over 2}{-c\over (2\pi)^5}\int_0^1 dz {z}\int
\Spaa{\la|d\la}\Spbb{\W\la|d\W\la} \int dt \delta\left( {z
K^2\over\Spab{\la|K|\W\la}} -  t \right)\left( {
K^2\over\Spab{\la|K|\W\la}}\right)^2 \nn &=&{\pi\over 2}{c K^2\over
2 (2\pi)^5},\eea
Comparing both results we find immediately $c=\pi/2$.

\subsection{The example with $f=(L_1+L_3)^2$}

First take the spinor method.
Notice that $s=(L_1+L_3)^2=(K-L_2)^2$, which can be used to simplify the calculation significantly,
since the integrations over $L_1$ and $L_3$ are the same as that of the pure phase space integration.
Then we have
\bea I_3^s(s;K) &=& \int
{d^4L_2\over(2\pi)^3}\delta^+(L_2^2)(K^2-2KL_2)\cdot{1\over 8\pi}\nn
&=& {-\pi^2\over 2(2\pi)^5}\int_0^1 dz {z(1-z)\over 2}\int
\Spaa{\la|d\la}\Spbb{\W\la|d\W\la}{(K^2)^3\over\Spab{\la|K|\W\la}^2}
= { \pi^2(K^2)^2\over 24(2\pi)^5}.\eea
%

Similar to equation (\ref{i3m1}), one has $s=(L_1+L_3)^2= Q^2 (1-x_2)$ (as
given in \cite{Field}), thus
\bea I_3^{m}(s;K)={Q^2\over 16 (2\pi)^3} \int_0^1 dx_1
\int_{1-x_1}^1 dx_2Q^2(1-x_2)={(Q^2)^2\over 96 (2\pi)^3}.\eea
%

\subsection{The example with $f=s^2 t$}

The integration is
\bea I_3^s(s^2t;K) =\prod_{i=1,2,3}\int\frac{
d^4L_i}{(2\pi^3)}\delta^+(L_i^2)(2\pi)^4\delta^4(K-L_3-L_2-L_1)(2L_1L_3)^2\cdot
2L_2L_3.\eea
This calculation is a little more complicated.
However, the basic steps are the same as those for $f=s$.
First, integrate over $L_1$ with the delta function of the
energy-momentum conservation;
then, integrate over $L_2$ with the remaining delta functions in spinor coordinates;
finally, integrate over $ L_3$ according to eq (\ref{I-phase}).
The process is generic.

For illustrations, here we write out the main steps
\bea I_3^s(s^2t;K)&=&\frac{1}{(2\pi)^5} \int
d^4L_3\delta^+(L_3^2)\int d^4L_2\delta^+(L_2^2)
\delta^+((K-L_3-L_2)^2)\times \nn && {} \left((2KL_3)^2\cdot
2L_2L_3-2(2KL_3)(2L_2L_3)^2+(2L_2L_3)^3 \right)\nn
&=&\frac{1}{(2\pi)^5}{\pi\over 2} \int d^4L_3\delta^+(L_3^2)
\left({1\over2}-{2\over3}+{1\over4}\right)(2KL_3)^3\nn
&=&-\frac{1}{(2\pi)^5} {\pi^2\over 24}\int_0^1 dz {z^4\over 2}\int
\Spaa{\la|d\la}\Spbb{\W\la|d\W\la}
{(K^2)^5\over\Spab{\la|K|\W\la}^2}=\frac{\pi^2(K^2)^4}{240(2\pi)^5}.\eea

Using the momentum method {\cite{Field} we have
\bea I_3^{m}(s^2t;K)={Q^2\over 16 (2\pi)^3} \int_0^1 dx_1
\int_{1-x_1}^1 dx_2(Q^2)^2(1-x_2)^2Q^2(1-x_1)={(Q^2)^4\over 960
(2\pi)^3},\eea
which is identical to $I_3^s(s^2t;K)$.

\section{Example three: four out-going particles}

In this section we show how to deal with multiple out-going particles recursively.
To be concrete, we will focus on the two-four process.
Generalization to cases with arbitrary $n$-outgoing particles will be straightforward.

\subsection{The pure phase space integration}
Using $I_3^s(1;K)$ as obtained in the previous section and the recursion relation, we have
\bean I_4^{s}(1;K) &=&\frac{\pi}{(2\pi)^3} \int
d^4L_4\delta^+(L_4^2)I_3^s(1;K-L_4)=
\frac{1}{4(2\pi)^8} {\pi^2\over 2}\int d^4L_4\delta^+(L_4^2)(K-L_4)^2\\
&=&\frac{1}{4(2\pi)^8}{\pi^2\over 2}\left(K^2\int
d^4L_4\delta^+(L_4^2)-\int d^4L_4\delta^+(L_4^2)2KL_4\right).\eean
Integrating out these two terms separately, one has
\bea I_4^{s}(1;K) = \frac{(K^2)^2}{768(2\pi)^5}.\eea
which can compare  with the calculation using momentum integration
method. Other method, like using the Lorentz invariant variables as
$x_{ij}\equiv L_i\cdot L_j$, can also be used to do the calculation.
However, unlike the situation with only three $L_i$, thing becomes
more complicated.

\subsection{Two examples of non-trivial $f$}

The first example is $f_1=s$, and the second example is
$f_2=(L_1+L_2)^2(L_2+L_3)^2(L_3+L_4)^2(L_4+L_1)^2
=(2L_1L_2)(2L_2L_3)(2L_3L_4)(2L_4L_1)$, which is more complicated.
Here the integrations will be performed with the spinor method. The
momentum method can be used to make comparisons. For $f_1$, the
evaluation is trivial and will not be presented. For $f_2$, the
calculation is a little involved and will be deferred to the
Appendix A.

For the $f_1=(L_1+L_2)^2$, we have
\bea I_4^{s}(s;K) =\frac{\pi}{(2\pi)^3} \int d^4L_4 \delta^+(L_4^2)
I_3^{s}(s;K-L_4)=\frac{1}{(2\pi)^3} {\pi^2\over 2}\int d^4L_4
\delta^+(L_4^2){((K-L_4)^2)^2\over12(2\pi)^5},\eea
by using the result for $I_3^s(s;K)$ obtained in the last section.
The last integration is easy to do.
Following the prescribed procedure, we get
\bea I_4^{s}(s;K)=\frac{(K^2)^3}{4608(2\pi)^5}.\eea

Now we turn to the function
$f_2=(2L_1L_2)(2L_2L_3)(2L_3L_4)(2L_4L_1)$. The factor $(2\pi)^{-8}$
and the Jacobi factor will be dropped for simplicity in the
following. The first step is almost the same as that of the momentum
method (notice for $I_2$, the $K$ is in fact the $K-L_3-L_4$),
\bea \label{I_2^s(f;K)} I_2^s(f_2;K) &=& \int
d^4L_2\delta^+(L_2^2)\int
d^4L_1\delta^+(L_1^2)\delta^4(K-L_2-L_1)(2L_1L_2)(2L_2L_3)(2L_3L_4)(2L_4L_1)\nn
&=& \int d^4L_2\delta^+(L_2^2)\delta^+((K-L_2)^2)\nn && \times
\left((2L_3L_4)(2KL_4)(2KL_2)(2L_2L_3)-(2L_3L_4)(2KL_2)(2L_2L_3)(2L_2L_4)\right).\eea
The first term is simple and it is ${(K^2)(2L_3L_4)(2KL_3)(2KL_4)/2}$.

The second term is a little bit more complicated.
To proceed, we convert the momentum measure into the spinor measure,
and perform the $t$-integral with the $\delta$-function:
\bea \label{4-s}&& \int d^4L_2\delta^+(L_2^2)\delta^+((K-L_2)^2)
(2L_3L_4)(2KL_2)(2L_2L_3)(2L_2L_4)\nn &=&
(2L_3L_4)\int\Spaa{\la|d\la}\Spbb{\W\la|d\W\la}{(K^2)^4\Spab{\la|L_3|\W\la}\Spab{\la|L_4|\W\la}
\over \Spab{\la|K|\W\la}^4}.\eea
Then, we  do an auxiliary integration as suggested in section 1.1:
\bea I_{aux} &=& \int\Spaa{\la|d\la}\Spbb{\W\la|d\W\la}
{(K^2)^4\Spab{\la|R|\W\la}^2\over \Spab{\la|K|\W\la}^4} \eea
with $R=xL_3+yL_4$. From eq. (\ref{res-I_{aux}}), one has
\bea I_{aux}&=& {1\over3}\left((K^2)(2KR)^2-(K^2)^2R^2\right).\eea
Next, pick the coefficient of the (2xy)-term in $I_{aux}$, which is
\bea{1\over3}(K^2)(2KL_3)(2KL_4)-{1\over6}(K^2)^2(2L_3L_4).\eea
Thus, the end result of (\ref{4-s}) is
\bea{1\over3}(K^2)(2KL_3)(2KL_4)(2L_3L_4)-{1\over6}(K^2)^2(2L_3L_4)^2.\eea
Combining the two terms in (\ref{I_2^s(f;K)}), we have
\bea I_2^s(f;K-L_3-L_4) &=&
{1\over6}(K^2)(2KL_3)(2KL_4)(2L_3L_4)+{1\over6}(K^2)^2(2L_3L_4)^2.\eea
There is no more difficulty in remaining steps.
After some elementary manipulations, we have
\bea I_3^{s}(f_2;K-L_4) &=& \int d^4L_3\delta^+(L_3^2)I_2^s(f_2;K-L_3-L_4)
={7\over 4320}(K^2)^3(2KL_4)^2,\\
I_4^{s}(f_2;K) &=& \int
d^4L_4\delta^+(L_4^2)I_3^{s,r}(f_2;K-L_4)={1\over 172800}(K^2)^6.
\eea
Putting all factors back, we get finally
\bea I_4^{s}(f_2;K) &=& {1\over 675 \times 2^{12}
(2\pi)^5}(K^2)^6.\eea
as to be checked with the calculations in the Appendix A.

\section{Conclusion}

In this paper, we have proposed to use the spinor integration method, which
is developed from the phase space integration of one-loop unitarity
cut, to do the real physical phase space integration for the total
cross section. This paper, as the initial construction of frame,
focus on the massless particles in pure 4D. We will discuss the
massive  and general D-dimension case in subsequent work.

Now, let us sum some salient points of our method. The first point
is the rewriting measure given in (\ref{I-phase}) by the
Faddeev-Popov trick. This new form enable us to do the spinor
integration method. Since the $z$-integration is always $[0,1]$ for
massless case, we can change the order of integration and leave all
$z$-integration at the end after we have performed all spinor
integrations. With $n$ out-going particles, there will be $(n-2)$
$z$-integration to do. The integrand will be the Lorentz invariant
expression of momenta of in-coming particles only. With two
particles $p_1, p_2$ we have only one nontrivial variables
$K^2=(p_1+p_2)^2$. For massive case, the integrand for $z$ will
depend on $K$ as well as various masses and thing will be more
complicated.

In comparison, there will be $(3n-4)$ integrations in general,
if we use the momentum integration method
\footnote{With a suitable choice of reference framework, we may reduce dimension
of integration somewhat further.}.
Furthermore,
the expressions are usually in component-form which is not manifest Lorentz invariant.
With more and more out-going particles,
it will also become difficult to specify the integrated regions and
separate angles variables and module variables.

One can also try to rewrite the measure using Lorentz invariant
variables. With $n$ $L_i$ there are ${n(n-1)\over 2}$ quantities
given by $x_{ij}=L_i\cdot L_j$. They are not independent to each
other in general, so the discussion of proper choice of subset as
well as the constraints among them becomes more and more tedious
with the increase of number of $L_i$.

Having shown the promise of our new method, the result in this paper
is not immediately to be useful for practical calculations. As we
have mentioned several times, we need to include massive outgoing
particles as well as general D-dimension. This can, in fact, be
accomplished by mimicking one-loop calculations, where one
generalizes pure 4D spinor integration to $(4-2\eps)$ dimension
spinor integration \cite{Anastasiou:2006jv, Anastasiou:2006gt}
although some technical difficulties need to be attacked. Work on
this is in progress.

{\bf Acknowledgement:} We would like to thank  C.~Anastasiou,
R.~Britto, and Z.~Kunszt for reading the draft. B.F would also like
to thank the hospitality of CUFE where the final part is done. The
work is funded by Qiu-Shi funding from Zhejiang University and
Chinese NSF funding under contract No.10875104, 10425525 and
10875103.

\appendix

\section{Using momentum method to integrate $f_2$ for four out-going particles}

In this appendix, we will present the integration with
$f_2=(2L_1L_2)(2L_2L_3)(2L_3L_4)(2L_4L_1)$ for four out-going
particles by using the momentum method.
It is to be compared with the spinor method presented in section 5.2.
For simplicity, the factor $(2\pi)^{-8}$ has been dropped, just as in section 5.2.
In the spirit of recursion, we first perform the integration
\bea \label{I_2^m(f;K)} I_2^m(f_2;K) &=& \int {d^3L_2 \over
2E_2}\int {d^3L_1 \over 2E_1}
\delta^4(K-L_2-L_1)(2L_1L_2)(2L_2L_3)(2L_3L_4)(2L_4L_1)\nn &=& \int
{d^3L_2 \over 2E_2}\delta^+((K-L_2)^2)\nn && \times
\left((2L_3L_4)(2KL_4)(2KL_2)(2L_2L_3)-(2L_3L_4)(2KL_2)(2L_2L_3)(2L_2L_4)\right).\eea
In the following, $K$ will be substituted by $K-L_3-L_4$.
The first term is simpler than the second. It is
\bea && \int {d^3L_2 \over
2E_2}\delta^+((K-L_2)^2)(2L_3L_4)(2KL_4)(2KL_2)(2L_2L_3)\nn
&=&
(2L_3L_4)(2KL_4) \int {d^3L_2 \over
2E_2}\delta^+(E^2-2EE_2)(2EE_2)\Big(2E_2E_3(1-z)\Big), \quad
(z\equiv\cos{\th_{23}})\nn
&=&
(2L_3L_4)(2KL_4)\cdot
{2\pi\over8}(E^2)(2EE_3)={2\pi\over8}(K^2)(2KL_4)(2KL_3)(2L_3L_4).\eea
In the last line, we have transformed from the component-form  to the Lorentz-invariant form,
such as $2EE_3\rightarrow2KL_3$, to make it suitable for the use of recursion relations.
Such manipulations will always be used in the following.

Next we deal with the second term (without the minus sign).
The difficulty is that the integrand depends on two angles,
$\th_{23}$ and $\th_{24}$, which cannot be integrated over at the same time.
However we can spilt the integral into two terms,
either of which has only one angle. To do this, we introduce two
auxiliary parameters $L_+=L_3+L_4$ and $L_-=L_3-L_4$. Then the
second term becomes
\bea && \int {d^3L_2 \over
2E_2}\delta^+((K-L_2)^2)(2L_3L_4)(2KL_2)(L_2(L_++L_-))(L_2(L_+-L_-))\nn
&=& (2L_3L_4) \int {d^3L_2 \over
2E_2}\delta^+((K-L_2)^2)(2KL_2)\left[(L_2L_+)^2-(L_2L_-)^2\right].\eea

Now we calculate the first term of the expression only.
The second term is almost the same as the first.
Set $L_+=(E_+,P_+\hat p_+),
L_-=(E_-,P_-\hat p_-)$, where $P_+$ and $P_-$ are the module of the
corresponding momentum vector. $\theta_{2+}$  is the angle between
$\vec l_2$ and $\hat p_+$, and $\theta_{2-}$ is that between $\vec
l_2$ and $\hat p_-$. Then the first term becomes
\bea \label{A} && (2L_3L_4)\int_0^{2\pi}d\varphi \int_{-1}^1dz_+
\int {E_2^2\over2E_2}dE_2
\delta^+(E^2-2EE_2)(2EE_2)(E_2E_+-E_2P_+z_+)^2,
\quad(z_+\equiv\cos\theta_{2+})\nn &=&
(2\pi)(2L_3L_4)\left({E^2\over16}(EE_+)^2+{E^4\over48}P_+^2\right).\eea
Also we have $L_+^2=E_+^2-P_+^2=(L_3+L_4)^2=2L_3L_4$, i.e.
$P_+^2=E_+^2-2L_3L_4$. Substituting it into eq (\ref{A}), we have
\bean
(2\pi)\left({1\over12}(K^2)(2L_3L_4)(KL_+)^2-{1\over48}(K^2)^2(2L_3L_4)^2\right).\eean
Similarly, the second term gives
\bean(2\pi)\left({1\over12}(K^2)(2L_3L_4)(KL_-)^2+{1\over48}(K^2)^2(2L_3L_4)^2\right).\eean
Combining them together, we get the result of the second term in
(\ref{I_2^m(f;K)}):
\bea(2\pi)\left({1\over12}(K^2)(2KL_4)(2KL_3)(2L_3L_4)-{1\over24}(K^2)^2(2L_3L_4)^2\right).\eea

Then
\bea I_2^m(f_2;K) &=& \left({2\pi\over8}-{2\pi\over
12}\right)(K^2)(2KL_4)(2KL_3)(2L_3L_4)+{2\pi\over24}(K^2)^2(2L_3L_4)^2\nn
&=&
{2\pi\over24}(K^2)(2KL_4)(2KL_3)(2L_3L_4)+{2\pi\over24}(K^2)^2(2L_3L_4)^2.\eea
The rest calculations are tedious but not difficult.
The end results are
\bea I_3^{m}(f_2;K) &=& \int {d^3L_3 \over 2E_3}I_2^m(f_2;K-L_3)={7\over34560}(2\pi)^2(K^2)^3(2KL_4)^2,\\
I_4^{m}(f_2;K) &=& \int {d^3L_4 \over
2E_4}I_3^{m}(f_2;K-L_4)={(2\pi)^3\over 2764800}(E^2)^6. \eea
Adding back the factor of $(2\pi)^{-8}$, we finally get
\bea I_4^{m}(f_2;K) &=& {1\over 2764800(2\pi)^5}(E^2)^6.\eea
which is the same as $I_4^{s}(f_2;K)$, as obtained in section 5.2.

\section{The spinor integration as integration in complex plane}

In this section, we will rewrite spinor integration as the integration
in the complex plane. The correctness of the
spinor integration method will be obvious. Furthermore, as complex
integration can be applied to any form of inputs in principal, so can the
spinor integration be. This is  very important
since for phase space integration, the input has singularities
other than those in the propagators in the one-loop calculation (see, for
example, \cite{Czakon:2008ii}). We will demonstrate it with a few
examples.

\subsection{Rewriting of spinor integration}

Since spinor $\la$ has only two components, we can expand it with
two independent spinors, for example, $\la_a, \la_b$ as
\bea |\la\rangle=|a\rangle+z|b\rangle, \quad
|\W\la]=|a]+\bar{z}|b]~.~~~~\label{S-exp-1}\eea
For real momentum, $\O z$ is the complex conjugation of $z$. Using
this expansion, we have $\Spaa{\la\ d\la}=\Spaa{a\ b}dz$,
$\Spbb{\W\la\ d\la}=\Spbb{a\ b}d\bar{z}$,
thus the measure is given by
\bea \int \Spaa{\la|d\la} \Spbb{\W\la|d\W\la}=
\Spaa{a|b}\Spbb{a|b}\int dzd \O z~.\eea
With  an arbitrary integrand  $G(\la,\W\la)$, using above
replacement we have
\bea \int \Spaa{\la|d\la} \Spbb{\W\la|d\W\la} G(\la,\W\la)=
\Spaa{a|b}\Spbb{a|b}\int dzd \O z G(z,\O z)~.\eea
In another word, the spinor integration is  a
two-dimensional integration over complex plane.

For integration in complex plane, there is an important formula
\cite{Complex-book}
\bea \oint_C (u dx +v dy)=\int\int_R \left( {\partial v\over
\partial x}-{\partial u\over
\partial y}\right) \eea
or in the form
\bea \oint_C g d z = -\int\int_R {\partial g\over \partial \O z}
dz\wedge d\O z~.~~~~~\Label{S-form-1} \eea
Now we see how to apply above formula with the simplest example:
\bea  & &  \int\Spaa{\la\   d\la}\Spbb{\W\la\ d\W\la}{1\over
\Spab{\la|K|\la}^2}=\int\int\Spaa{a\ b}\Spbb{a\ b}dzd\bar{z}{1\over
(\Spab{\la|K|a}+\bar{z}\Spab{\la|K|b})^2}\nonumber \\
&=&\int\int\Spaa{a\ b}\Spbb{a\ b}dzd\bar{z} {d\over d\O z}{1\over
\Spab{\la|K|b}}{-1\over
(\Spab{\la|K|a}+\bar{z}\Spab{\la|K|b})}\nn
& =& -\oint\Spaa{a\ b}\Spbb{a\ b}dz{1\over \Spab{\la|K|b}}{-1\over
(\Spab{\la|K|a}+\bar{z}\Spab{\la|K|b})}~.\eea
Using $\Spab{\la|K|b}=\Spab{a|K|b}+z\Spab{b|K|b}$ with  pole
location $z=-{\Spab{a|K|b}\over\Spab{b|K|b}}$. Taking residue of
pole we get ${-1\over K^2}$ (where we have neglected the $2\pi i$
factor).

It will be useful to compare above calculation with the one we did
in spinor integration. We have
\bea  \int\Spaa{\la\ d\la}\Spbb{\W\la\ d\W\la}{1\over
\Spab{\la|K|\la}^2}&=&\int\Spaa{\la\ d\la}\Spbb{d\W\la\
\partial_{\W\la}}{ \Spbb{\eta\ \W\la}\over \Spab{\la|K|\eta}\Spab{\la|K|\W\la}}~. \eea
Taking the residue of $\Spab{\la|K|\eta}$ we have ${-1\over K^2}$.
If we set $\eta=b$ and using
\bea d\O z {\partial \over \partial \O z}=
\Spbb{d\W\la|\partial_{\W\la}}~,~~~~\label{reverse} \eea
we will see explicitly the one-to-one correspondence between our
spinor integration method and complex integration method.

The power of (\ref{S-form-1}) is that it reduces the two-dimension
integration into pure algebraic calculation, i.e., reading out
residue by implicitly using
\bea  {\partial \over \partial \O z} {1\over
z-a}=\pi\delta^2(z-a)~.\eea
Spinor integration method has also used this property, which is
casted into the spinor formula and called the ``holomorphic anomaly"
\cite{Cachazo:2004by}.

It is obvious now that any spinor integration can be rewritten  as
complex plane integration, so its validity is obvious. However,
keeping the spinor form the expression will be much more compact and
also easier to read out residues.

\subsection{More examples}

The spinor method is developed for the one-loop calculation at
beginning, thus the most cases we have met are propagator-like
singularities, where  we have result (\ref{GETPOLE}). For other
kinds of singularities, we need similar results. This can be
obtained by first going to complex variables and then pulling back
to spinor variables. Using this technique we can do the spinor
integration for other kinds of singularities.  Let us demonstrate
this with a few examples.

First let us discuss
\bea & &  \int\Spaa{\la\ d\la}\Spbb{\la\ d\la}{1\over\
\Spab{\la|K|\la}\Spab{\la|R|\la}}\nn
 &=&\int\Spaa{a\ b}\Spbb{a\
b}dzd\bar{z}{1\over
(\Spab{\la|K|a}+\bar{z}\Spab{\la|K|b})(\Spab{\la|R|a}+\bar{z}\Spab{\la|R|b})}\nn
&=&\int\Spaa{a\ b}\Spbb{a\ b}dzd\bar{z}({\Spab{\la|K|b}\over
\Spab{\la|K|a}+\bar{z}\Spab{\la|K|b}}-{\Spab{\la|R|b}\over
\Spab{\la|R|a}+\bar{z}\Spab{\la|R|b}})\nn &&\times {1\over
\Spab{\la|K|b}\Spab{\la|R|a}-\Spab{\la|R|b}\Spab{\la|K|a}}\nn &=&
\int \Spaa{a\ b}d\bar{z}{\partial \over \partial
\bar{z}}(\ln({\Spab{\la|K|a}+\bar{z}\Spab{\la|K|b}})-\ln({\Spab{\la|R|a}+\bar{z}\Spab{\la|R|b}}))\nn
&&\times {1\over \Spaa{\la|KR|\la}} \eea
Using (\ref{reverse})  as well as
\bea \Spbb{\a|{\partial\over \partial \W\la}}
\Spbb{\b|\W\la}=\Spbb{\b|\a}\eea
above result  is equivalent to
\bea \Spbb{d\W\la\  \partial_{\W \la}}\left( {1\over
\Spaa{\la|K|R|\la}}\ln {\Spab{\la|K|\W\la}\over
\Spab{\la|R|\W\la}}\right)=\Spbb{\W\la\ d\W\la} {1\over
\Spab{\la|K|\W\la}\Spab{\la|R|\W\la}}\eea
Similarly we have
\bea \Spbb{d\W\la\  \partial_{\W \la}}\left( {1\over
\Spaa{\la|K|R|\la}}\sqrt{ {\Spab{\la|R|\W\la}\over
\Spab{\la|K|\W\la}}}\right)={-1\over 2}\Spbb{\W\la\ d\W\la} {1\over
\Spab{\la|K|\W\la}^{3\over 2}\Spab{\la|R|\W\la}^{1\over 2}} \eea
Using these results, we can do more examples:

{\bf Example 1:}
The first example is the square root form as
\bea I_1&=&\int
d^4L_1\delta^+(L_1^2)\delta^+((L_1-K)^2){1\over\sqrt{L_1\cdot R}}\nn
&=&\int \Spaa{\la\ d\la}\Spbb{\wt\la\ d\wt\la}\sqrt{2\over
K^2}{1\over \Spab{\la|K|\wt\la}^{3/2}\sqrt{\Spab{\la|R|\wt\la}}}\nn
&=&\int \Spaa{\la\ d\la}{2\sqrt{2\over K^2}\over
\Spaa{\la|K|R|\la}}\Spbb{d\W\la|{\partial\over \partial \W\la}}
{-\sqrt{\Spab{\la|R|\wt\la}}\over \sqrt{\Spab{\la|K|\wt\la}}}
\label{sqrt-root} \eea
Reading out residues we have
\bea I_1&= &{2 (\sqrt{2K\cdot R-\sqrt{\Delta}}+\sqrt{2K\cdot
R+\sqrt{\Delta}})\over K^2 \sqrt{\Delta}}\eea
where $\Delta= (2R\cdot K)^2-4R^2K^2$.\\

{\bf Example 2:}

The second example is the one with gram determinant. This is, in
fact, the one we will meet in one-loop calculation as well as in
tree-level after one external particle having been integrated out.
It represents the typical singularity behavior in the phase space
integration.
\bean I_2=\int d^4L\delta^+(L^2)\delta^+((L-K)^2)&=&\int
d^4L\delta^+(L^2)\delta^+((L-K)^2){1\over \det \left(
  \begin{array}{ccc}
    K_1^2 & K_1\cdot K_2 & K_1\cdot K_3 \\
    K_1\cdot K_2 & K_2^2 & K_3\cdot K_2 \\
    K_1\cdot K_3 & K_3\cdot K_2 & K_3^2 \\
  \end{array}
\right) }\nn
 &=&\int
{d^4L\delta^+(L^2)\delta^+((L-K)^2)\over 2(K\cdot R)(L\cdot
R)(K\cdot L)-R^2(K\cdot L)^2-K^2(L\cdot R)^2 }\nn \eean
where we have taken $K_1=K,K_2=R,K_3=L$. Doing standard manipulation
we get
\bea I_2&=&\int\Spaa{\la\ d\la}\Spbb{\la\ d\la}{-4\over K^2 }{1\over
2(K\cdot
R)\Spab{\la|K|\la}\Spab{\la|R|\la}-R^2\Spab{\la|K|\la}^2-K^2\Spab{\la|R|\la}^2}\nn
& = & \int\Spaa{\la\ d\la}\Spbb{\la\ d\la}{-4\over K^2
}{1\over\Spab{\ell|x_1 R+ y_1 K|\ell}\Spab{\ell|x_2 R+ y_2 K|\ell}}
~~~~\Label{I2-fac}\eea
where
\bean y_2=1,~~~y_1=-R^2,~~~x_1=K\cdot R+\sqrt{ (K\cdot R)^2- K^2
R^2},~~~x_2={-K\cdot R+\sqrt{ (K\cdot R)^2- K^2 R^2}\over R^2}\eean
Now this is the familiar case given by (\ref{I-B}) and we have the
final result
\bea I_2&=& {-4\over K^2}\int_0^1 d\alpha {1\over
((1-\alpha)(x_1R+y_1K)+\alpha (x_2R+y_2 K))^2}\nn
& = & {1\over K^2 ((K\cdot R)^2-K^2 R^2)}\int_0^1 d\alpha {1\over
\alpha(\alpha-1)} \eea
which is divergent with $\a$-integration. This example is, in fact,
another typical example of divergence at the boundary of phase
space.

It is worth to emphasize that the factorization (\ref{I2-fac})  we
have done is the general method we need to use for phase space
integration in more complicated situations.


\end{document}